\documentclass[conference]{IEEEtran}
\IEEEoverridecommandlockouts
\usepackage{cite}
\usepackage{amsmath,amssymb,amsfonts}
\usepackage{algorithmic}
\usepackage{graphicx}
\usepackage{graphics}
\usepackage{textcomp}
\usepackage{calrsfs}
\usepackage{lipsum}
\usepackage{float}
\usepackage{subcaption}
\usepackage[table,xcdraw]{xcolor}
\usepackage{booktabs}
\usepackage{multirow}
\DeclareMathAlphabet{\pazocal}{OMS}{zplm}{m}{n}
\newcommand{\Lb}{\pazocal{L}}
\usepackage{xcolor}
\usepackage{hyperref}
\def\BibTeX{{\rm B\kern-.05em{\sc i\kern-.025em b}\kern-.08em
    T\kern-.1667em\lower.7ex\hbox{E}\kern-.125emX}}
\usepackage{adjustbox} %Francisco

\begin{document}

%CAMBIAR TITULO
\title{Multi-Scale Structural-aware Exposure Correction for Endoscopic Imaging\\
%{\footnotesize \textsuperscript{*}Note: Sub-titles are not captured in Xplore and
%should not be used}
\thanks{Identify applicable funding agency here. If none, delete this.}
}

\author{\IEEEauthorblockN{Axel García-Vega}
\IEEEauthorblockA{\textit{School of Engineering and Sciences} \\
\textit{Tecnologico de Monterrey}\\
Nuevo Leon, Mexico \\
0000-0002-1504-2428}
\and
\IEEEauthorblockN{Ricardo Espinosa}
\IEEEauthorblockA{\textit{Faculty of Engineering} \\
\textit{Universidad Panamericana}\\
Aguascalientes, Mexico \\
email address or ORCID}
\and
\IEEEauthorblockN{Gilberto Ochoa-Ruiz}
\IEEEauthorblockA{\textit{School of Engineering and Sciences} \\
\textit{Tecnologico de Monterrey}\\
Nuevo Leon, Mexico \\
email address or ORCID}
\and
\IEEEauthorblockN{Luis Falcón-Morales}
\IEEEauthorblockA{\textit{School of Engineering and Sciences} \\
\textit{Tecnologico de Monterrey}\\
Nuevo Leon, Mexico\\
email address or ORCID}
\and
\IEEEauthorblockN{Dominique Lamarque}
\IEEEauthorblockA{\textit{dept. name of organization (of Aff.)} \\
\textit{name of organization (of Aff.)}\\
City, Country \\
email address or ORCID}
\and
\IEEEauthorblockN{Christian Daul}
\IEEEauthorblockA{\textit{dept. name of organization (of Aff.)} \\
\textit{name of organization (of Aff.)}\\
City, Country \\
email address or ORCID}
}

\author{Axel García-Vega$^{1}$, Ricardo Espinosa$^{2,4}$, Luis Ramírez-Guzmán$^{1}$, Thomas Bazin$^{3}$, Luis Falcón-Morales$^{1}$, \\ Gilberto Ochoa-Ruiz$^{1}$, Dominique Lamarque$^{3}$ and Christian Daul$^{2}$,\\

\thanks{$^{1}$School of Engineering and Sciences, Tecnologico de Monterrey, Mexico}
\thanks{$^{2}$CRAN (UMR 7039, Université de Lorraine and CNRS), Nancy, France}%
\thanks{$^{3}$H\^opital Ambroise Paré (AP-HP), Boulogne-Billancourt France}%
\thanks{$^{4}$Universidad Panamericana, Aguascalientes, Mexico}%
\thanks{*Contacts: gilberto.ochoa@tec.mx, christian.daul@univ-lorraine.fr}
}
\maketitle
\begin{abstract}
Endoscopy is the most widely used imaging technique for the diagnosis of cancerous lesions in hollow organs. However, endoscopic images are often affected by illumination artefacts: image parts may be over- or underexposed according to the light source pose and the tissue orientation. These artifacts have a strong negative impact on the performance of computer vision or AI-based diagnosis tools. Although endoscopic image enhancement methods are greatly required, little effort has been devoted to over- and under-exposition enhancement in real-time. This contribution presents an extension to the objective function of LMSPEC, a method originally introduced to enhance images from natural scenes. It is used here for the exposure correction in endoscopic imaging and the preservation of structural information. To the best of our knowledge, this contribution is the first one that addresses the enhancement of endoscopic images using deep learning (DL) methods. Tested on the Endo4IE dataset, the proposed implementation has yielded a significant improvement over LMSPEC reaching a SSIM increase of 4.40\% and 4.21\% for over- and underexposed images, respectively.
\end{abstract}

\section{Introduction}\label{Intro}

%General Intro (good part of endoscopy) Main areas and successes of endoscopy in recent years, the role computer vision

Endoscopy plays a central role in minimally invasive surgery or for carrying out examinations in hollow organs, such as the colon or the stomach.  In recent years, computer aided endoscopy has become an important area of research. In particular, Computer Vision (CV) has the potential of becoming an essential tool for assisting endoscopists in various tasks\cite{juanca2022, Zenteno2022, Martinez2000}. 

\begin{figure}
    \centering
    \includegraphics[width=0.45\textwidth]{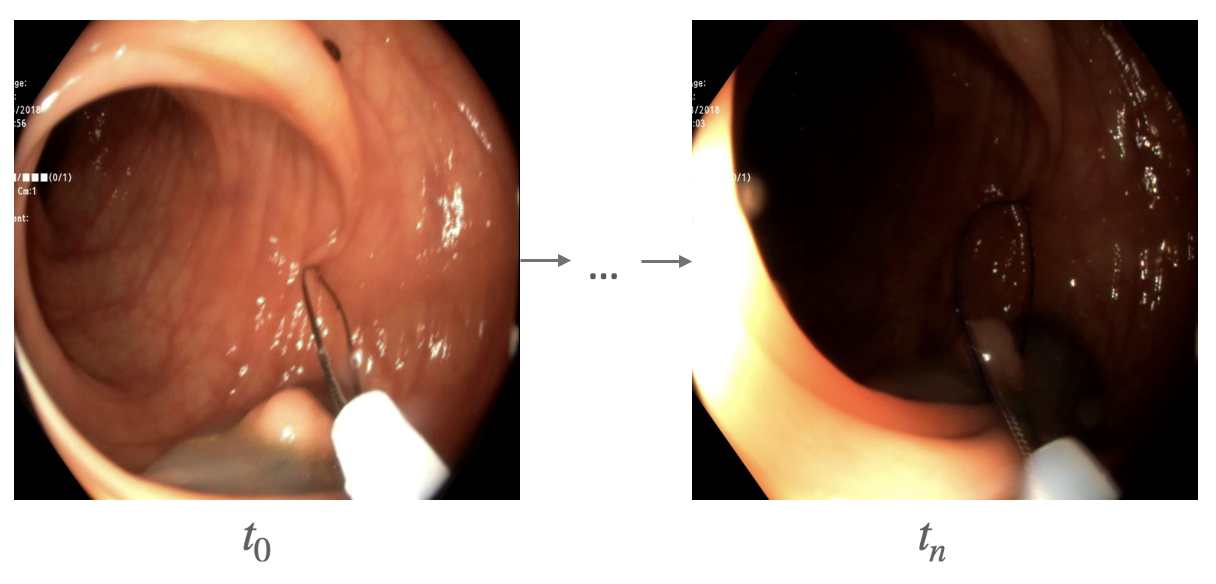}
    \caption{Strong illumination change example in almost consecutive frames of a colonoscopic image sequence. (a) This image was acquired in appropriate lighting conditions. (b) Few frames later, the image is overexposed in its lower left region and underexposed in the remaining frame part.}
    \label{sequential}
\end{figure}

However, a major hurdle that most of these CV methods must face is related to the uncontrolled and highly changing illumination conditions in endoscopic scenes. Figure \ref{sequential} shows two colonoscopic images in which strong illumination changes are visible.  Such uncontrolled lighting affects the robustness of Computer-Aided Detection (CADe) and Diagnosis (CADx). The performance of techniques for recovering extended surfaces of hollow organs (such as SLAM \cite{chen2018slam} or Structure for Motion \cite{phan2020optical}) is also affected by uncontrolled lighting. These strong photo-metric variations are due to non-optimal light source poses, moist surfaces, and occlusions that lead to under- or overexposed video-frames parts \cite{ma2019real}. 

Therefore, any improvements of endoscopic image content quality could considerably boost the efficiency of CV- and AI-based CAD tools. In this regard, various challenges have been proposed in conferences to foster the development of algorithms which can be bench-marked in terms of generalization capabilities. One such challenge is the Endoscopic Artifact Detection challenge (EAD, \cite{ali2021deep}), with includes various types of endoscopic artefacts for developing novel image preprocessing algorithms.

The results obtained by numerous methods in the EAD challenge have shown that image enhancement (IE) algorithms are of high interest for improving the robustness and generalization capabilities of endoscopic image preprocessing techniques. This contribution focuses  on the exposure correction in white light endoscopy. It is noticeable that this issue has only been partially addressed in the IE field, as most methods (see \cite{afifi2021learning})  were dedicated to the correction of either under- or over-exposed images, but did not deal with both effects occurring concurrently. Contrary to images of natural scenes, in endoscopy imaging it is common that both types of non-optimal exposures simultaneously affect frames. Thus, a preprocessing algorithm should be able to detect and correct in real-time all types of inappropriate exposures.

García-Vega et al. proposed a paired ``normal-exposed'' image dataset \cite{garcia2022novel, Endo4IE} to assess the ability of machine learning-based methods to correct the effects of non-optimal lighting conditions. However, the need for both accurate and real-time IE techniques highlighted the shortcomings of most current methods. Nonetheless, the LMSPEC deep-learning (DL) method proposed by Afifi et al. \cite{afifi2021learning} outperformed other models in terms of accuracy and inference time, whilst obtaining a satisfactory enhancement performance. However, in some images, LMPSEC introduced undesired textural and color artifacts, which could lead, for instance, to false diagnoses in endoscopy or errors in automated methods.

This contribution shows how to alleviate the loss of texture and color information during  the exposure correction process by introducing a structural similarity-aware extension to the overall loss function of the LMSPEC pipeline. This modification allows to preserve fine-textured details. The results show that is possible to achieve this goal both for under- and over-exposures, while maintaining a relatively low inference time.

The rest of the paper is organized as follows. Section \ref{SofA} gives an overview of the works relating to endoscopic image enhancement. Section \ref{Materials} discusses the dataset used to perform the experiments presented in this contribution,  as well as the metrics used to evaluate the performance of the solution introduced in Sections \ref{DL-model} (DL-model) and \ref{ExpSetUp} (DL-model tuning). Section \ref{ExpRes} gives, through a set of ablation studies, quantitative and qualitative comparisons between the different configurations of the proposed DL-model.  Finally, Section \ref{Conclusions} globally discusses the results and outlines perspectives.

\section{State of the Art}\label{SofA}

% REALLY SUPERFICIAL, REDUNDANT AND REPETITIVE  
%Restoration and image enhancement (IE) is the process of reconstructing an correcting images with the main purpose of improving the visual quality of images, while providing reliable information for subsequent computer vision (CV) tasks.  For the purpose of this contribution, we will use the IE term to refer to the correction of under and over exposed sections of an image frame. %In photography, these techniques have been traditionally applied for correction of exposure effects as underexposure and overexposure.

In the past, IE methods based on different approaches have been proposed, such as methods using histogram equalization \cite{wang2008flattest} or Retinex theory-based models \cite{LOL}. 
More recently, frameworks based on machine learning models have emerged. In these approaches, the IE mapping is learned instead of computed, and DL models have excelled at this task since its inception. Although these DL-methods have shown great promise, they still suffer from various shortcomings. First and foremost, most of the methods in the existing literature can enhance only under- or over-exposed images, but cannot simultaneously perform both tasks with efficiency. Thus, \cite{guo2020zero}, \cite{LOL} and \cite{wang2022low} are examples of methods that can be successfully applied in under-exposed images, taking advantage of deep-learning networks and the Retinex assumption. Another drawback of the most existing models is that they take several seconds to enhance the quality of a single image. Such processing times make applications are inappropriate when  real-time performance is required, such as for CADe and CADx tools for endoscopy or 3D reconstruction pipelines for colonoscopy. \cite{ceron2021assessing}

Nonetheless, some recent methods have been proposed to address the enhancement of both under- and overexposed images with inference times which are near to real-time. For instance, Afifi et al. \cite{afifi2021learning} proposed a novel coarse-to-fine DL-based technique which is trained in a fully supervised manner. The method divides the IE task in two sub-problems, color (including brightness or exposure) enhancement and detail enhancement. For bidirectional exposure correction, a Laplacian pyramid (derived from a Gauss pyramid) is used for performing a multi-resolution decomposition method. The Laplacian pyramid decomposition allows to enhance the color and detail (texture) information sequentially.

In a prospective study, Garcia et al. \cite{garcia2022novel} compared various image enhancement methods (among them were also the LMSPEC method), using a recent dataset containing synthetically generated under- and over-exposed endoscopic frames which are paired with their non-corrupted ground truth image. The authors assessed the capabilities of the different IE methods to enhance the quality of endoscopic images, while maintaining a high degree of fidelity. To do so, the texture quality was quantified in terms of peak signal to noise ratio (PSNR) and structural similarity index measure (SSIM), as well as subjectively graded by human evaluator. The inference times of each model was also measured. In this study, LMPSPEC demonstrated an astounding performance for both types of exposure artifacts, while attaining an almost real-time performance. However, the IE-model introduced, on the one hand, some artifacts that removed high frequency content (texture details) from the enhanced images, and led, on the other hand, on other undesired color artifacts. 

The main lesson of the tests conducted in \cite{garcia2022novel}, was in observation that, even if the LMSPEC method is (quantitatively and qualitatively) the most appropriate when both under- and over-exposures must be corrected in images, it also introduces some noticeable high-frequency artifacts (textures are altered). The aim of this contribution is to improve the LMSPEC method in terms of texture preservation.

\section{Materials}\label{Materials}

\subsection{Dataset}\label{Dataset}

%\textcolor{red}{DELETE ALL THIS, FOCUS ON THE DATA SET ITSELF:
%In endoscopic examinations, it is very common to come across with exposure errors due to light reflexions in the inner walls of the hollow organs. For instance, when the tip of the endoscope  (which has a light) points to folds, these structures reflect the light, provoking overexposure, and whereas, an underexposed region at the other end of the frame can appear. This can be a big problem for physicians since they are fully responsible of detecting anomalies, blood, or even polyps, that could be translated in warning cases. Currently, methods for enhancing exposure errors need paired data, i.e., corrupted frames and their respective ground-truth (i.e., non-corrupted or clean image). For instance, for natural images the LOL or MIT-Adobe FiveK datasets, which contain common real life images have been proposed. These paired datasets allow researchers to both train and evaluate their models making use of standardized ground-truth images. To the best of our knowledge, in endoscopic domain,  there are not publicly available in the literature. This is due to the difficulty of producing the same frame with and without exposures errors, naturally or with the help of a photo editing expert. Hence, our work aimed to create, through the use of GANs, a paired dataset of real images without any exposure error and the same image with an exposure error, that is a corrupted frame and its ground-truth.} 

The dataset used in this contribution is a combination of three different existing datasets (EAD \cite{AliEAD2020}, EDD \cite{ali2021deep} and HyperKVisir \cite{Borgli2020}) using the procedure described in \cite{garcia2022novel}. The authors used image-to-image translation to take unmodified endoscopic frames and generated frames with over- and underexposure artefacts.  For implementing this task, they used the CycleGAN architecture \cite{zhu2017unpaired} since the main issue to tackle was the lack of paired data (CycleGAN is an efficient method for working with unpaired data). 

This dataset is composed of three different types of images: i) 2216 unmodified (acquired) endoscopic frames (without exposure errors) that act as ground truth data, ii) 1,231 synthetically overexposed frames, and iii) 985 synthetically underexposed frames. Every ground truth image in sub-dataset 1 is associated with its either over- or underexposed synthetic version, belonging to sub-datasets 2 and 3, respectively. Both paired sub-datasets (sub-datasets 1+2 and sub-dataset 1+3) were split as follows: 70\% , 27\% and 3\% of the images were used for the training,  test, and validation steps, respectively. 

\subsection{Metrics}\label{Metrics}
Two standard full-reference metrics were used to compare different IE methods: PSNR) and SSIM. Both metrics can evaluate   globally over the image  the quality of the obtained results. 

 \begin{figure*}[!ht]
    \centering
    \includegraphics[scale=0.48]{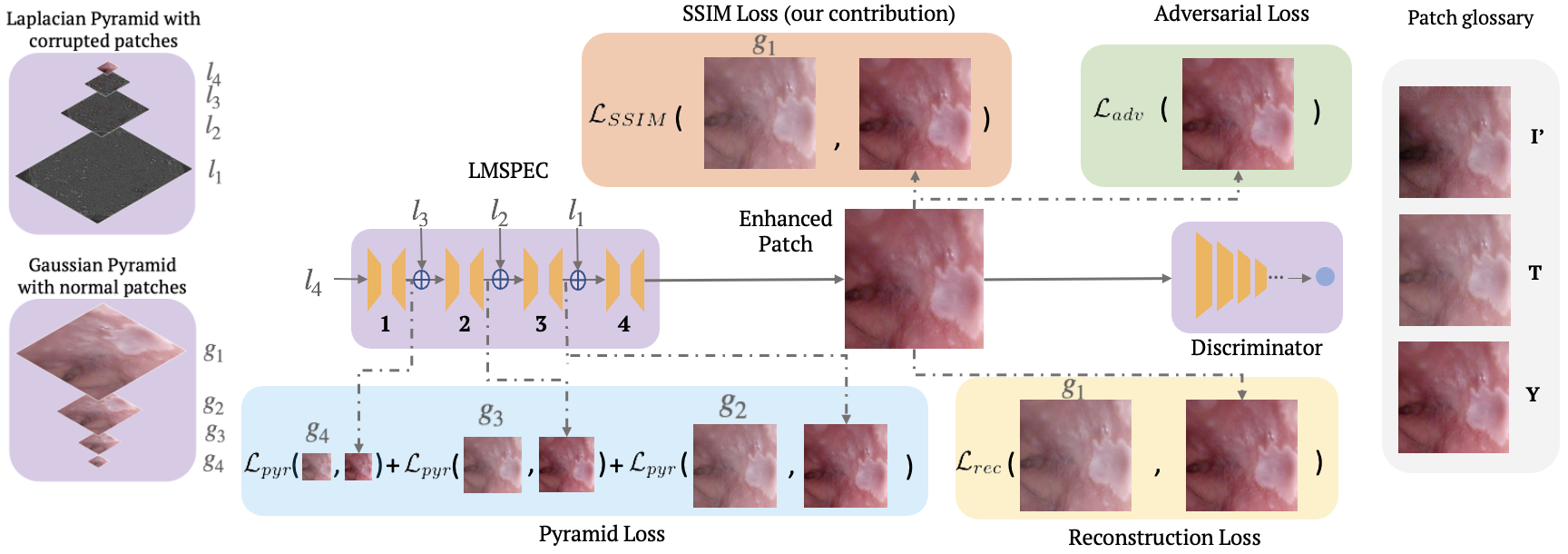}
    \caption{DL-mode. On the left: Laplacian pyramid decomposition over patches \textbf{I'} with exposure artefacts and Gaussian pyramid decomposition over ground truth patches \textbf{T}. On the right: $\Lb_{pyr}$ is computed with the up-sampled output from sub-networks \textbf{1},\textbf{2} and \textbf{3}, whereas $\Lb_{rec}$, $\Lb_{SSIM}$ and $\Lb_{adv}$ are computed with the final up-sampled output \textbf{Y} from the sub-network \textbf{4}. In addition, when the discriminator network is enabled, it is simultaneously trained with the final output and its respective ground truth.}
    \label{loss_diagram}
\end{figure*}

\section{Proposed DL Method}\label{DL-model}

Given a poorly exposed input image \textbf{I} acquired under white-light, the proposed method (depicted in Figure \ref{loss_diagram}) aims to predict an output image \textbf{Y} being a version of \textbf{I} with no exposure errors. As in LMSPEC, the color and detail errors of \textbf{I} are sequentially corrected. Basically,  a multi-resolution representation of \textbf{I} is given by a pyramidal Laplacian decomposition derived from a pyramidal Gaussian decomposition of ground truth \textbf{T}.

The heart of the method is based on the original implementation of LMSPEC, which randomly extracts $n$ small patches $I'_{1},...,I'_{n}$ from \textbf{I} and decomposes each patch into two components: i) a four-level Gaussian Pyramid (GP) and then ii) a four-level Laplacian Pyramid (LP). This LP can be seen as a set of frames with different frequency levels, $LP=\{l_{1}, l_{2}, l_{3}, l_{4}\}$, where $l_{1}$ and $l_{4}$ contain the high- and low frequency components, respectively. This LP decomposition is carried out to feed four U-Net-like sub-nets in a cascade configuration with sub-images with different levels of detail. Each sub-net is used to extract relevant features from the image and to carry out a reconstruction of each $l_{i}$ input in reverse order, as shown in the LMSPEC  block in Fig. \ref{loss_diagram}.

%The decomposition of each patch into four pyramids was experimentally determined as the configurations by the LMSPEC authors. 

%\begin{figure}[!ht]
%    \centering
%    \includegraphics[scale=0.43]{figures/pyramids.png}
%    \caption{Four-level Laplacian pyramid from exposed random extracted patches and Gaussian pyramid from the ground truth equivalent patches.}
%    \label{pyramids}
%\end{figure}

Processing input image \textbf{I} in this manner permits to independently deal with each sub-net output and compute Pyramid Loss $\Lb_{pyr}$. This loss is the weighted sum of the four \textit{$\Lb_{1}$} losses, one for each last three LP level predictions. Thus, in order to compute $\Lb_{pyr}$, the target for each level is given by the Gaussian pyramid $GP$ of the patch extracted from ground truth $T$. For $GP=\{g_{1}, g_{2}, g_{3}, g_{4}\}$, $\Lb_{pyr}$ is computed as follows:

\begin{equation}
    \Lb_{pyr} = \sum\limits_{i=2}^4 2^{i-2}\Lb_{1}(\hat{g_{i}}, \hat{l_{i}}).
\label{eq1}
\end{equation}

The value of $\Lb_{rec}$ is also based on the $\Lb_{1}$ loss, which measures the pixel-wise error between the prediction and the ground truth patches $T_{j}$ as shown in (\ref{eq2}), where $j$ is the j-th patch extracted from the frame.

\vspace{-4mm}

\begin{equation}
    \Lb_{rec} = \Lb_{1}(T_{j}, Y_{j}),
\label{eq2}
\end{equation}

The last sub-net makes the final prediction $Y_{j}$, which is used to compute three of the losses: i) a Reconstruction Loss $\Lb_{rec}$ ii) an Adversarial Loss $\Lb_{adv}$ and iii) a Structural Similarity Loss $\Lb_{SSIM}$. In \cite{shao2022self}, Shao et. al combined the $\Lb_{1}$ and $\Lb_{SSIM}$ losses to improve the enhancement results in comparison to a single use of the $\Lb_{1}$ loss. As discussed above, both the $\Lb_{pyr}$ and $\Lb_{rec}$ losses are based on $\Lb_{1}$ and thus, by adding the $\Lb_{SSIM}$ loss (see Eq. \ref{eq3}) to the overall training objective, enforces the model to learns from the pixel distribution in the ground truth patch, thus leading to a model with more consistent outputs without increasing the inference time of the method. The attractive characteristic of the SSIM loss is the fact that it has been proved to be successful when dealing with complex illumination changes \cite{bian2019unsupervised}. This fact enabled the proposed approach to improve the results of the original LMSPEC implementation.
\begin{equation}
    \Lb_{SSIM} =   (1-SSIM(T_{j}, Y_{j})/2
    \label{eq3}
\end{equation}

%\begin{equation}
%    SSIM(x, y) = \frac{(2\mu_{x} \mu_{y} + c_{1})(2\sigma_{xy} + c_{2})}{(\mu_{x}^{2}+\mu_{y}^{2}+c_{1})(\sigma_{x}^{2}+\sigma_{y}^{2}+c_{2})}.
%    \label{eq4}
%\end{equation}

For preserving realism, LMSPEC integrates a Discriminator, which takes {\bf Y} as input and returns a scalar score that indicates how realistic the image looks like. This block is trained along the main network and is used for computing an adversarial loss $\Lb_{adv}$ shown in (\ref{eq5}). 

In this contribution, the loss function for optimizing the discriminator is the same as in \cite{goodfellow2020generative}:
\vspace{-1mm}
\begin{equation}
    \Lb_{adv} = -3hwn\log(S(D(Y_{j}))),
    \label{eq5}
\end{equation}

where $n$ is the number of pyramid levels ($4$ in this paper) and $S(D(Y_{j}))$ is the sigmoid function applied to the D value of the final prediction or the generated image. 
The complete loss function is then computed as follows:
\begin{equation}
    \Lb = \alpha \Lb_{pyr} + \beta \Lb_{rec} + \gamma \Lb_{SSIM} + \delta \Lb_{adv},
\label{eqfinale}
\end{equation}

where, $\alpha$, $\beta$, $\gamma$ and $\delta$ are regularization weights. Figure \ref{loss_diagram} shows how each single loss was computed through out the entire pipeline.

%\begin{table}[!b]
%\centering
%\caption{Configuration of regularization parameters for the ablation study. The best configuration is in the row highlighted.}
%\begin{tabular}{ccccc}
%\hline
%Config. & $\alpha$ & $\beta$ & $\gamma$ & $\delta$ \\ \hline
%1      & 1.0      & 1.0     & 1.0      & 1.0      \\
%2      & 1.0      & 0.25    & 0.25     & 0.25     \\
%3      & 0.25     & 1.0     & 0.25     & 0.25     \\
%\rowcolor[HTML]{EFEFEF} 
%4      & 0.25     & 0.25    & 1.0      & 0.25     \\
%5      & 0.25     & 0.25    & 0.25     & 1.0      \\ \hline
%\end{tabular}
%\label{tab1}
%\end{table}

%\vspace{-5mm}

\section{Experimental Setup and Model tuning }\label{ExpSetUp}

The DL-model parameter tuning was carried out as follows. First, an ablation study was carried out to determine the appropriate values of the regularization parameters ($\alpha$, $\beta$, $\delta$ and $\gamma$). Then, we fine-tuned the best of these configurations. Furthermore, a three-fold training stage was performed with a single underexposed (UE) dataset, a single overexposed (OE) dataset and a combined over-underexposed (C) dataset (as in \cite{afifi2021learning}). The best model from the ablation study was given by following parameter configuration: $\alpha=\beta=\delta= 0.25$ and $\gamma=1.0$. This setting gives a strong importance to the SSIM term, which allows to preserve texture details. Moreover, this model (\textit{Baseline}) and LMSPEC were initially trained with original hyper-parameters as shown in upper part of Table \ref{tab1}. Since the input type used in this contribution is different from the one in the original implementation, the hyper-parameters were tuned to maximize the performance on each training sub-dataset (UE, OE and C). The best hyper-parameters after training with each sub-dataset yielded three fine-tuned separated models, as seen in the lower part of Table \ref{tab1}. It is worth noticing that each training was done in two phases as follows: first the trained used 128 pixel square patches, then the weights were transferred as initialization of the second training phase with 256 pixel patches. For this second training phase, the discriminator was enabled at certain discriminator starting epoch (DSE) specified in configurations in Table \ref{tab1}.

%\vspace{-2mm}

%For training our modified architecture, we carried out the followed procedure. First, we trained a set of baseline models through and ablation study (to be dicussed in the section \ref{ExpRes}). The best performed configuration was then trained using patches with size 128x128 in order to maximize the number of training samples from the original dataset. This configuration was trained for 50 epochs, using a batch size of 32, with a learning re of $10^{-4}$ for the generator, with a half-drop learning rate each 20 epochs.
%Afterwards, the model with this weight initialization was further trained on 256x256 patches for 40 epochs to refine the model. The discriminator blocks must be trained in intervals (i.e., enabled) in order to discern between generated and ground truth effectively, starting time at epoch 20, and batch size of 8 a learning rate of $10^{-5}$. In both cases, we used transposed 2D convolutions.

% Please add the following required packages to your document preamble:
% \usepackage{multirow}
% \usepackage[table,xcdraw]{xcolor}
% If you use beamer only pass "xcolor=table" option, i.e. \documentclass[xcolor=table]{beamer}
\begin{table}[]
\centering
\caption{Hyper-parameter configurations. Phase 1 (128 pixels  patches) in white rows, phase 2 (256 pixels patches) in gray.}
\begin{adjustbox}{width=0.95\columnwidth,center}
\begin{tabular}{@{}ccccccc@{}}
\toprule
Method                  & Training Set                & Epochs                     & DSE                        & BS                         & $lr_{G}$                                & $lr_{D}$                                \\ \midrule
LMSPEC \cite{afifi2021learning}  &    & 40 & -  & 32 & $10^{-4}$ & $10^{-5}$ \\
Baseline & \multirow{-2}{*}{UE, OE, C} & \cellcolor[HTML]{EFEFEF} 30 & \cellcolor[HTML]{EFEFEF}15 & \cellcolor[HTML]{EFEFEF}8  & \cellcolor[HTML]{EFEFEF}$10^{-4}$ & \cellcolor[HTML]{EFEFEF}$10^{-5}$ \\ \midrule
 &   & 50 & -  & 32 & $10^{-4}$        &$10^{-5}$        \\
                        & \multirow{-2}{*}{UE}        & \cellcolor[HTML]{EFEFEF}40 & \cellcolor[HTML]{EFEFEF}20 & \cellcolor[HTML]{EFEFEF}8  & \cellcolor[HTML]{EFEFEF}$10^{-4}$        & \cellcolor[HTML]{EFEFEF}$10^{-5}$        \\[0.15cm] 
                        &                             & 40 & -  & 64 & $2\times10^{-4}$ & $2\times10^{-5}$ \\
                        & \multirow{-2}{*}{OE}        & \cellcolor[HTML]{EFEFEF}30 & \cellcolor[HTML]{EFEFEF}15 & \cellcolor[HTML]{EFEFEF}32 & \cellcolor[HTML]{EFEFEF}$2\times10^{-4}$ & \cellcolor[HTML]{EFEFEF}$2\times10^{-5}$ \\[0.15cm]
                        &   & 50 & -  & 32 & $10^{-4}$        & $10^{-5}$        \\
\multirow{-6}{*}{Best models*} & \multirow{-2}{*}{C}         & \cellcolor[HTML]{EFEFEF}40 & \cellcolor[HTML]{EFEFEF}20 & \cellcolor[HTML]{EFEFEF}8  & \cellcolor[HTML]{EFEFEF}$10^{-4}$        & \cellcolor[HTML]{EFEFEF}$10^{-5}$        \\ \bottomrule 
\multicolumn{7}{l}{\footnotesize *Fine-tuned models for each sub-dataset. DSE: discriminator starting epoch.} \\
%\multicolumn{3}{l}{\footnotesize DSE: discriminator starting epoch.  DSE: discriminator starting epoch.}\\
\multicolumn{7}{l}{\footnotesize BS: batch size. lr: learning rate.}\\
%\multicolumn{3}{l}{\footnotesize lr: learning rate.}\\
\end{tabular}
\end{adjustbox}
\label{tab1}
\end{table}

%We carried out an ablation study to find the best hyper-parameters for the losses by grid search. In the first experiment, we did not apply regularization, then in the following experiments we took a pivot and a regularization of $0.25$ over the other loss functions in order to see which one outperforms.%, the five experiments are shown in Table \ref{tab1}.

%The best configuration of the model regarding to our ablation study was given by $\alpha=\beta=\delta= 0.25$ and with no regularization $\gamma=1.0$ over our proposed loss. From Table \ref{tab2} it can be observed that, in terms of SSIM metric, the performance our model is greater than the original LMSPEC implementation for overexposure and underexposure correction. This result was expected as it gives more weight to the SSIM loss. This model was finally chose for further hyper-parameter tuning (i.e., the two stages number of epochs, discriminator training start time, batch size, learning rate, and upsampling technique). 

\section{Results and Discussion}\label{ExpRes}

\subsection{Quantitative Results}\label{quan}

%In \cite{garcia2022novel} several models were compared with LMSPEC obtaining a significant performance difference against traditional methods, for both under and over exposure correction.

Table \ref{tab2} gives an overview of of the results for the inference phase, which has been carried out over each exposure type, i.e., the UE and OE models were tested over under- and overexposure patch sets respectively, whereas  the model C was tested over both (separated) test sets.

The results of the proposed model are compared with those of the baseline LMSPEC model. Table \ref{tab2} shows that the proposed method outperforms LMSPEC best model (either for separated sets or combined) outperforms LMSPEC in terms of SSIM by 4.40\% and 4.21\% for over-exposed and under-exposed images, respectively. 
Therefore, also note that best performance of our proposed method was given by training our proposed model plus fine-tuning with separated datasets.

%which might be explained from the over-smoothing that LMPSEC introduces in the predicted image due to the lack of the SSIM loss. 

\begin{table}[!t]
\centering
\caption{Quantitative results on th Endo4IE dataset \cite{garcia2022novel}. White rows: independent-data training. Light gray: combined-data training as in \cite{afifi2021learning}. Highest criterion values are in bold. }
\label{tab:my-table}
\begin{tabular}{@{}ccccc@{}}
\toprule
                         & \multicolumn{2}{c}{Overexposure} & \multicolumn{2}{c}{Underexposure}                                                \\ \midrule
Method  & PSNR$\uparrow$  & SSIM$\uparrow$  & PSNR$\uparrow$  & SSIM$\uparrow$  \\ \midrule
                         & 21.846          & 0.744          & \textbf{24.204} & 0.757          \\
\multirow{-2}{*}{LMSPEC \cite{afifi2021learning}} & \cellcolor[HTML]{EFEFEF}22.286          & \cellcolor[HTML]{EFEFEF}0.772          & \cellcolor[HTML]{EFEFEF}23.064          & \cellcolor[HTML]{EFEFEF}0.760          \\[0.2cm]
                         & 22.633          & 0.799          & 23.720          & 0.783          \\
\multirow{-2}{*}{Baseline}   & \cellcolor[HTML]{EFEFEF}22.442          & \cellcolor[HTML]{EFEFEF}0.795          & \cellcolor[HTML]{EFEFEF}22.877          & \cellcolor[HTML]{EFEFEF}0.786          \\ [0.2cm]
  & \textbf{23.139} & \textbf{0.806} & 24.201          & \textbf{0.792} \\
\multirow{-2}{*}{Baseline*}  & \cellcolor[HTML]{EFEFEF}22.704 & \cellcolor[HTML]{EFEFEF} 0.801 & \cellcolor[HTML]{EFEFEF} 23.229 & \cellcolor[HTML]{EFEFEF} 0.786 \\ \bottomrule
\multicolumn{4}{l}{\footnotesize *Proposed model + fine-tuned model.} \\
\end{tabular}
\label{tab2}
\end{table}

%\vspace{-3mm}

\subsection{Qualitative Results}\label{qua}

\begin{figure}[!t]
    \centering
    \includegraphics[scale=0.45]{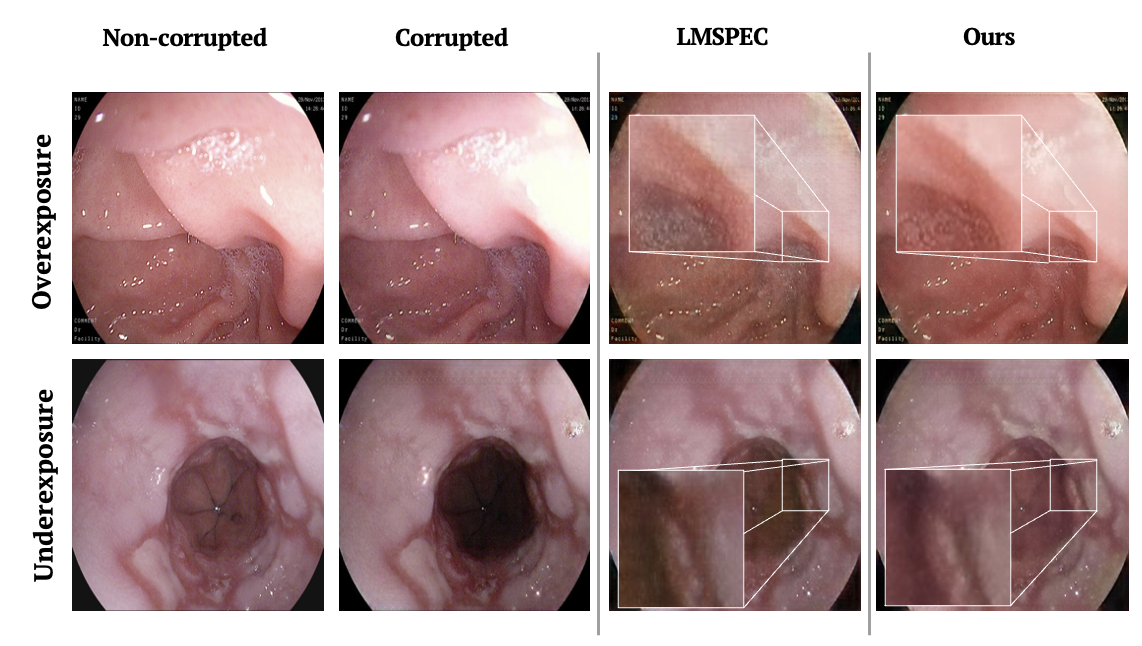}
    \caption{Visual assessment of the exposure correction and structure preservation.  The structural enhancement is perceptible in the zoomed areas. The complete images include less artifacts.}
    \label{visual_results}
\end{figure}

Figure \ref{visual_results} shows a qualitative comparison for a couple of frames from the Endo4IE \cite{Endo4IE} dataset. From the zoomed areas in the third and fourth columns (images enhanced by LMSPEC and the proposed method, respectively) it can be observed that the proposed method is able to produce a much more reliable prediction in comparison to the ground truth (first column), both for over- and underexposed frames (second column). However, a slight change in hue is introduced by both methods. This issue requires further investigation. % We plan to improve this aspect in future work  with the introduction of histogram-based loss for improving the color preservation after the enhancement procedure. 

\section{Conclusions and Future Work}\label{Conclusions}

%In this paper, we proposed an extension of the DL-based LMSPEC model for carrying out exposure correction of under and over exposed frames. %We made use of a tailored dataset (Endo4IE) containing a paired dataset for assessing reference-based image enhancement methods. 

It was shown that the proposed extension of LMSPEC, in the form of an extra loss term for preserving texture details in exposure corrected images has yielded satisfactory results in the Endo4IE dataset: the experiments show a boost in terms of quantitative metrics and a qualitative assessment has shown that the method produces more realistic images. However, some improvements are still possible: i) although the model makes use of 7 million parameters, we have been able to attain only a 8 FPS throughput (high inference time), and ii) the model sometimes produces images with a slight shift in hue. The last issue can probably be addressed by enforcing color preservation via an additional loss, while the former issue requires improvements in the model design.

%Nonetheless, we consider that our results are encouraging and a larger dataset and the used of perceptual metric could foster the research in the exposure enhancement of endoscopic images. We also plan to integrate our method in a 3D reconstruction pipeline to assess the gain that can be attained by our using our enhacenment method.

\section*{Acknowledgments}

The authors wish to thank the AI Hub and the CIIOT at Tecnologico de Monterrey for their support for carrying the experiments reported in this paper in their NVIDIA's DGX computer. We also wish thank CONACYT for the master scholarship for Carlos Axel Garcia Vega at Tecnologico de Monterrey.
\bibliographystyle{splncs04}
\bibliography{references}

\end{document}